\documentstyle[preprint,prl,aps]{revtex}
\begin{document}

\draft

\title{Disorder Induced Phase Transition in a
Random Quantum Antiferromagnet}

\author{Anders W. Sandvik\cite{aboakademi} and Marco Veki\'c \cite{davis}}
\address{Department of Physics, University of California,
Santa Barbara, CA 93106}

\date{\today}

\maketitle

\begin{abstract}

A two-dimensional Heisenberg model with random antiferromagnetic
nearest-neighbor  exchange is studied using quantum Monte Carlo techniques.
As the strength of the randomness is increased, the system undergoes
a transition from an antiferromagnetically ordered ground state to a
gapless disordered state. The finite-size scaling of the staggered
structure factor and susceptibility is consistent with
a dynamic exponent $z = 2$.

\end{abstract}


\vfill\eject

The copper-oxygen layers of the parent compounds of the high-T$_c$
superconductors are good physical realizations of the two-dimensional
(2D) antiferromagnetic Heisenberg model \cite{undoped}. Recently,
significant progress has been made
in the theory of clean 2D antiferromagnets with short-range interactions
\cite{chakravarty,critical}. The nonlinear $\sigma$-model (nl$\sigma$m)
in $2+1$ dimensions is believed to describe their long wavelength physics.
Chakravarty {\it et al.} \cite{chakravarty} concluded that the ground state
of this field-theory can be either ordered or disordered, depending on the
coupling constant $g$, which depends on the interactions and the magnitude
of the spin. There is convincing numerical evidence that the $S={1\over 2}$
Heisenberg model with nearest-neighbor interactions on a square lattice
is on the ordered side of the phase diagram, with a sublattice magnetization
$m \approx 0.3$ \cite{heisenberg1}, which is close to the ordered moment
observed in the cuprates \cite{undoped}.

The microscopic mechanism leading to the destruction of the antiferromagnetic
long-range order upon doping the cuprates is still not understood.
In particular, in spite of some experimental support \cite{imai,sokol}, it
is not clear whether the nl$\sigma$m indeed describes the phase transition,
as the dopants are expected to induce strong randomness in the spin-spin
interactions before
the charge carriers become mobile \cite{aharony}. It is therefore important
to investigate  how various types of quenched disorder affect the long-range
order of the 2D Heisenberg model, and how the nl$\sigma$m description
\cite{chakravarty,critical} is altered by randomness.

In this Letter we present quantum Monte Carlo results for the $S={1\over 2}$
random exchange model
\begin{equation}
\hat H = \sum\limits_{\langle i,j\rangle} J_{ij} \vec S_i \cdot
\vec S_j ,
\label{hamiltonian}
\end{equation}
where ${\langle i,j\rangle}$ is a pair of nearest-neighbor sites on
a 2D square lattice. The couplings $J_{ij}$ take two values,
$J_{ij} = J(1 \pm \Delta)$, at random, with a probability $p$ for $1+\Delta$
and $1-p$ for $1-\Delta$. We consider only the case $\Delta < 1$, i.e.
all couplings are antiferromagnetic and the system is non-frustrated.
Although the hamiltonian (\ref{hamiltonian}) does not exactly represent
the kind of disorder present in the cuprates, where one would
expect light doping to cause random frustrated interactions \cite{aharony}
(corresponding to $\Delta > 1$), we expect it to be of relevance in gaining
understanding of the effects of randomness in quantum antiferromagnets.
In 2D an order-disorder transition can only take place at zero
temperature, where the critical behavior is governed by quantum fluctuations.
The opportunity to numerically study a disorder-driven quantum phase
transition is a further motivation for this work.

We study the behavior of the hamiltonian (\ref{hamiltonian}) in the
$(p,\Delta)$ plane. For $\Delta \to 0$ the clean 2D Heisenberg model
is recovered independently of $p$, and the system is hence ordered at
$T=0$. For $\Delta \to 1$
(but $\Delta \not=1$)  both limits $p\to 0$ and $p\to 1$ correspond
to the 2D Heisenberg model with dilute bond-impurities. Thus the
system should be ordered in these regimes as well. As $p$ is increased
from $0$ there will be an increasing fraction of singlets forming at isolated
strong bonds in a background of weakly coupled spins. We argue that at a lower
critical concentration $p=p_{c1}$, this leads to an order--disorder
transition, in analogy with order--disorder transitions due to singlet
formation in clean quantum antiferromagnets, such as the 2-layer
Heisenberg model \cite{twolayer} and various other dimerized models
\cite{dimers}. As $p$ is increased further, there must be another transition to
an ordered state at $p=p_{c2}$, as the strong bonds start to dominate and the
weak bonds  effectively become impurities in a background of strongly
coupled spins.
As $\Delta$ is lowered the tendency to singlet formation diminishes, and
one would expect the range $[p_{c1}(\Delta),p_{c2}(\Delta)]$ to become
smaller and eventually vanish at some $\Delta = \Delta_{min}$ \cite{delta1}.
Below we present numerical results supporting this picture,
which is illustrated by the phase diagram outlined in Fig. 1. The
solid circles at the phase boundary are results of our quantum Monte Carlo
simulations, which are discussed below.

We have used a modification of Handscomb's quantum Monte Carlo
technique \cite{method,exact}, and averaged over 50-300 realizations of the
random couplings in order to obtain results useful for extrapolation to
the thermodynamic limit. We have studied systems of $L^2$ spins with periodic
boundary conditions. In order to obtain ground state results
for $L=4-10$, we have carried out simulations at inverse temperatures
$\beta=J/T$ as large as 128, which for these system sizes is enough for
all calculated quantities to have saturated to their $T=0$ value.
A theorem by Lieb and Mattis \cite{lieb} guarantees
that the ground state of a finite system with
an even number of spins is a singlet, as long as all couplings are
antiferromagnetic. We have therefore restricted the simulations to
the subspace with zero magnetization ($\sum_i S^z_i =0$). We have also
studied lattices with $L=32$ at higher temperatures. In these simulations
Monte Carlo moves changing the total magnetization were included.

The sublattice magnetization $m$ for a finite system can be defined
according to
\begin{equation}
m^2 = 3S(\pi,\pi)/ L^2,
\end{equation}
where $S(\pi,\pi)$ is the staggered structure factor
\begin{equation}
S(\pi,\pi) = {1\over L^2} \sum\limits_{j,k}
\hbox{e}^{i\vec\pi \cdot (\vec r_j - \vec r_k)}
\langle S^z_jS^z_k \rangle .
\end{equation}
For the clean 2D Heisenberg model, spin-wave theory gives the leading size
dependence of $m^2$ as \cite{huse}
\begin{equation}
m^2(L) = m^2(\infty) + kL^{-1}.
\end{equation}
In Fig. 2 we show results for $m^2$ at a strong-bond
concentration $p=1/4$ for system sizes $L=4,6,8,10$, and various
values of the disorder strength $\Delta$. An approximately linear dependence
on $1/L$ is seen for $\Delta < 0.8$. We therefore fit straight lines to
those points and extrapolate to obtain the sublattice magnetization
for the infinite systems. For $\Delta=0$ we
obtain $m=0.276 \pm 0.004$, which is slightly lower \cite{magnet} than
the spin-wave result $m = 0.303$ \cite{huse}. For $\Delta =0.70$ the
extrapolated $m^2$ is zero within statistical errors, and this should
therefore be close to the critical disorder strength for $p=1/4$. For
$\Delta > \Delta_c$ the scaling with system size must change to
$(1/L)^2$ for large $L$, as $S(\pi,\pi)$ saturates. We have also
performed simulations at $p=1/8$ and $p=1/2$, and obtained
$\Delta_c \approx 0.75$ and $\Delta_c \approx 0.80$, respectively.
These points are shown as the solid circles in Fig. 1.
The rest of the phase boundary is outlined schematically. The result
for $\Delta_c(p=1/2)$ indicates that the critical concentration
$p_{c2}$ as $\Delta \to 1$ is larger than the percolation
threshold \cite{delta1}.

For a given $p$, with $p_{c1} < p < p_{c2}$, as $\Delta \to \Delta_c$ the
spatial correlation length
$\xi_r$ diverges as $\delta^{-\nu}$, where $\delta=|\Delta - \Delta_c|$.
The correlation length in imaginary time, $\xi_\tau$, diverges as
$\delta^{-z\nu}$, where $z$ is the dynamic exponent \cite{hertz}.
With $\Delta > 0$ Lorentz-invariance is broken, and one expects $z \not=1$.
The dynamic exponent can be determined by comparing the size-dependence
of the staggered structure factor $S(\pi,\pi)$ and the staggered
susceptibility $\chi (\pi ,\pi)$. For a zero-temperature quantum phase
transition the exponent $\eta$ for the algebraic decay of the spatial
correlation function $C(r)$ is defined by \cite{fisher}
\begin{equation}
C(r) \to {1\over r^{2-D+z+\eta}}, \quad r \to \infty,
\end{equation}
where $D$ is the spatial dimensionality. The staggered structure factor
therefore diverges as $\delta^{-\nu (2-z-\eta)}$.
The staggered susceptibility is given by
\begin{equation}
\chi(\pi,\pi) = {1\over L^2} \sum\limits_{j,k}
\hbox{e}^{i\vec\pi \cdot (\vec r_j - \vec r_k)}
\int\limits_0^\beta d\tau \langle S^z_j (\tau) S^z_k (0) \rangle ,
\end{equation}
and diverges as $\delta^{-\nu (2-\eta)}$.

Finite-size scaling theory \cite{barber} gives the size dependence
of $S(\pi ,\pi)$ and $\chi (\pi,\pi)$ at the critical point:
\begin{mathletters}
\begin{eqnarray}
S(\pi ,\pi) && \sim L^{2-z-\eta}, \\
\chi(\pi ,\pi) && \sim L^{2-\eta} .
\end{eqnarray}
\label{sizedep}
\end{mathletters}
Hence, if $S(\pi,\pi) \sim L^{\gamma_S}$ and $\chi(\pi,\pi)
\sim L^{\gamma_\chi}$, the dynamic exponent is given by
$z = \gamma_\chi - \gamma_S$. In Fig. 3, $\ln[S(\pi ,\pi)]$ and
$\ln[\chi(\pi ,\pi)]$ are graphed versus $\ln(L)$ for two points
which we estimate to be close to the phase boundary in the $(p,\Delta)$-plane.
Least-squares fits of straight lines give the slopes
$\gamma_s = 1.01 \pm 0.01$, $\gamma_\chi = 2.88 \pm 0.09$
for $p=1/4,\Delta=0.7$ and $\gamma_s = 1.00 \pm 0.02$,
$\gamma_\chi = 3.12 \pm 0.15$ for $p=1/2,\Delta=0.8$. Hence we have
two independent estimates for the dynamic exponent; $z=1.87 \pm 0.10$
and $z=2.12 \pm 0.15$. The indicated errors are of course only statistical.
There are also errors, which we believe smaller than the statistical
ones, due to the points investigated not being exactly on the phase
boundary.

The value of $z$ has consequences for the behavior of the uniform
susceptibiliy $\chi_u$ at the transition point. The hyperscaling theory
developed by Fisher {\it et al.} gives \cite{fisher}
\begin{equation}
\chi_u = {\beta\over N} \sum\limits_{i,j} \langle S^z_iS^z_j\rangle
\sim \delta^{\nu(D-z)} .
\label{uniform}
\end{equation}
Hence, depending on $z$, the uniform susceptibility diverges, remains
finite or vanishes at the critical point. For ``dirty bosons'',
Fisher {\it et al.} argued that the total compressibility
(which corresponds to $\chi_u$) remains finite, and $z=D$ \cite{fisher}.
The spin model considered here can be mapped onto a hard-core bose system
with particle-hole and $SU(2)$ symmetries. The additional symmetries might in
principle change the universality class from the
one of the systems considered in Ref. \cite{fisher}. Our results for
the dynamic exponent are, however, consistent with $z=2=D$. The
results displayed in Figs. 2 and 3 indicate that $\eta$ is close to $-1$,
which satisfies the bound $\eta < 0$ \cite{fisher}.
In principle, one can use finite-size scaling for $\Delta > \Delta_c$ to
obtain the correlation length exponent $\nu$. We do not presently
have sufficient data for a precise estimate, but the bound
$\nu \ge 2/D=1$ \cite{fisher,nubound} appears to be satisfied.

We have calculated the uniform susceptibility for $L=32$ systems at
temperatures $T/J=0.2-1.0$. Results for $p=1/2$ and various $\Delta$ are
displayed in Fig. 4. The disorder clearly enhances the low-temperature
susceptibility. In view of the fact that $\chi_u$ is non-zero at $T=0$ for the
clean 2D Heisenberg model, it is unlikely that the susceptibility of the
random systems vanishes as $T\to 0$. According to Eq. (\ref{uniform}),
this implies that $z \ge 2$, consistent with the above estimates from
finite-size scaling. If $z=2$, $\chi_u$ approaches a constant as $T\to 0$
at $\Delta=\Delta_c$. In 1D, random exchange leads to a low-temperature
divergence of the uniform susceptibility \cite{1dpapers}. This is also
predicted in higher dimensions for systems with longer-range interactions
\cite{semicond}. The natural interpretation of this behavior is that some
of the spins are effectively isolated from the rest of the system due to
their local environment of strongly coupled spins. One would expect this
behavior in the disordered phase  of the model considered here as well,
but not on the phase boundary if indeed $z=2$.

A non-zero $T=0$ uniform susceptibility implies that the disordered phase
is gapless. In order to further investigate the spectrum, we
have calculated the
imaginary-time correlation function
$C(\tau) = {1/ N}\sum_i \langle S^z_i (\tau) S^z_i(0)\rangle$,
and used the maximum entropy analytic continuation procedure \cite{maxent}
to obtain the real-time wave-vector integrated dynamic structure factor
$S(\omega) = {1/ N} \sum_q S(q,\omega)$. We have calculated $S(\omega)$ for
both clean and random systems. In addition, in order to test the method,
we have studied the case where the strong bonds are arranged in a
regular staggered pattern such that every spin belongs to a pair connected
by a strong bond ($p=1/4$). In this case one expects a gap for $\Delta$
larger than a critical value. Fig. 5 shows low-temperature
results for $L=10$ systems. For the staggered and random cases $\Delta=0.8$,
and the ground states of the corresponding infinite systems are disordered.
The staggered system exhibits a clear gap; $S(\omega)$
is a narrow peak centered at $\omega/J \approx 1+\Delta$, corresponding to
the energy required for a singlet-triplet excitation of a spin pair
connected by a strong bond. For the clean system there
is a broad maximum around $\omega/J \approx 2$, and a narrow peak close to
$\omega =0$. In the thermodynamic limit, long wavelength fluctuations of
the order parameter lead to a $\delta$-function peak at $\omega=0$. This
peak is here shifted to a non-zero energy by the small gap present in the
finite system. For the random case the peak at $\omega = 0$ is due to
localized, gapless excitations. There is also more weight at low energies than
for the clean system, which we associate with excitations involving primarily
the weak bonds.

It would clearly be interesting to study the properties of a random
quantum antiferromagnet with a gap. Two coupled layers, each described by
the hamiltonain (\ref{hamiltonian}), should have both gapped and gapless
disordered phases, depending on $p$, $\Delta$ and the (non-random)
inter-layer coupling. Work on this model is in progress.

We are grateful to M.P.A. Fisher for an enlightening discussion on
scaling theory. We also thank H. Monien, D. Scalapino
and N. Trivedi for helpful conversations. Most numerical results were
obtained by severe abuse of a farm of HP-715 work stations at the ECE
Department at UC Davis. We also acknowledge the Texas National
Research Commission for an allocation of computer time through
Grants RGFY9166 and RGFY9266. This work is supported by the Department
of Energy under Grant No. DE-FG03-85ER45197 (A.W.S.) and the
National Science Foundation under Grant No. DMR92-06023 (M.V.).

\begin{figure} FIG. 1.
Proposed phase diagram of the random exchange model (\ref{hamiltonian})
in the $(p,\Delta)$-plane. The solid circles are Monte Carlo estimates
of transition points between the antiferromagnetic (AF) and gapless
disordered (GD) phases. The curve is a schematic outline of the rest of the
phase boundary.
\end{figure}

\begin{figure} FIG. 2.
The sublattice magnetization squared versus the inverse system size
at a strong-bond concentration $p=1/4$ and $\Delta=0$ (solid circles),
$\Delta=0.5$ (open circles), $\Delta=0.7$ (solid squares),
$\Delta=0.8$ (open squares), and $\Delta=0.9$ (solid triangles). Where
not shown, statistical errors are smaller than the size of the symbols.
The dashed and solid lines are fits to the $\Delta=0$ and $\Delta =0.7$
data, respectively.
\end{figure}

\begin{figure} FIG. 3.
Finite-size scaling of the staggered structure factor (open circles) and
the staggered susceptibility (solid circles) for $p=1/2,\Delta=0.8$
(top) and $p=1/4,\Delta=0.7$ (bottom). The straight lines are least-squares
fits to the points.
\end{figure}

\begin{figure} FIG. 4.
The uniform spin susceptibility versus temperature for systems of
size $32\times 32$, at a strong-bond concentration $p=1/2$. Solid circles
are for $\Delta=0$, open circles for $\Delta=0.5$, solid squares for
$\Delta=0.7$, open squares for $\Delta=0.8$, and solid triangles for
$\Delta=0.9$.
\end{figure}

\begin{figure} FIG. 5.
Wave-vector integrated dynamic structure factors for $L=10$. The dashed
curve is for a clean system at $\beta=32$, the solid
curve for a random system with $p=1/4$, $\Delta=0.8$ at $\beta=64$, and the
dotted curve for a staggered strong-bond pattern with $\Delta=0.8$.
\end{figure}


\begin{references}

\bibitem[(a)]{aboakademi} On leave from Department of Physics, {\AA}bo Akademi
University, {\AA}bo, Finland.

\bibitem[(b)]{davis} Permanent address:
Department of Physics, University of California, Davis, CA 95616

\bibitem{undoped}
G. Aeppli {\it et al.}, Phys. Rev. Lett. {\bf 62}, 2052 (1989);
J. M. Tranquada {\it et al.}, Phys. Rev. B {\bf 40}, 4503 (1989);
N. Pyka {\it et al.}, Z. Phys. B {\bf 82}, 177 (1991);
S.M. Hayden {\it et al.}, Phys. Rev. Lett. {\bf 67}, 3622 (1991).

\bibitem{chakravarty}
S. Chakravarty, B.I. Halperin, and D.R. Nelson,
Phys. Rev. Lett. {\bf 60}, 1057 (1988); Phys. Rev. B {\bf 39}, 2344 (1989).

\bibitem{critical}
S. Sachdev and J. Ye, Phys. Rev. Lett. {\bf 69}, 2411 (1992);
A. Chubukov and S. Sachdev, Phys. Rev. Lett. {\bf 71}, 169 (1993).

\bibitem{heisenberg1}
J.D. Reger and A.P. Young, Phys. Rev. B {\bf 37}, 5987 (1988);
M. Gross {\it et al.} Phys. Rev. B {\bf 39}, 2484 (1989);
K. Runge, Phys. Rev. B {\bf 45}, 7229 (1992);
U.J. Wiese and H.P. Ying Z. Phys. B {\bf 93}, 147 (1994).

\bibitem{imai}
T. Imai {\it et al.} Phys. Rev. Lett. {\bf 70}, 1002 (1993);
{\bf 71}, 1254 (1993).

\bibitem{sokol}
A. Sokol and D. Pines, Phys. Rev. Lett., {\bf 71}, 2813 (1993),

\bibitem{aharony}
A. Aharony {\it et al.} Phys. Rev. Lett. {\bf 60}, 1330 (1988).

\bibitem{twolayer}
K. Hida, J. Phys. Soc. Jpn. {\bf 61}, 1013 (1992);
A.J. Millis and H. Monien, Phys. Rev. Lett. {\bf 70}, 2810 (1993);
A.W. Sandvik and D.J. Scalapino, Phys. Rev. Lett. {\bf 72}, 2777 (1994).

\bibitem{dimers}
N. Katoh and M. Imada, J. Phys. Soc. Jpn. {\bf 62}, 3728 (1993);
S.R. White, R.M. Noack and D.J. Scalapino, preprint (1994).

\bibitem{delta1}
Exactly at $\Delta = 1$ the system cannot be ordered for $p$ smaller
than the percolation threshold $p_p = {1\over 2}$. Quantum fluctuations
may destroy the long-range order even for $p > p_p$. One would expect the
critical probability $p_{c2}(\Delta \to 1)$ to coincide with the critical
value of $p$ at $\Delta = 1$.

\bibitem{method}
A.W. Sandvik and J. Kurkij\"arvi, Phys. Rev. B {\bf 43}, 5950 (1991);
A.W. Sandvik, J. Phys. A {\bf 25}, 3667 (1992).

\bibitem{exact}
In this approach there are no $\Delta\tau$ errors of the type present
in world-line methods; J. E. Hirsch {\it et al.}, Phys. Rev. B
{\bf 26}, 5033 (1982).

\bibitem{lieb}
E.H. Lieb and D.C. Mattis, J. Math. Phys. {\bf 3}, 749 (1962).

\bibitem{magnet}
The numerical values of $m^2$ for $L=4-10$ agree well with Runge's
Green's function Monte Carlo results (Ref. \cite{heisenberg1}). He icluded
a quadratic term in the fit and obtained a slightly larger extrapolated
magnetization.

\bibitem{huse}
D.A. Huse, Phys. Rev. B {\bf 37}, 2380 (1988).

\bibitem{hertz}
J.A. Hertz, Phys. Rev. B {\bf 14}, 1165 (1976).

\bibitem{fisher}
M.P.A. Fisher {\it et al.}, Phys. Rev. B {\bf 40}, 546 (1989).

\bibitem{barber}
See, for example,
M.N. Barber, in {\it Phase transitions and critical phenomena,  vol. 8.},
edited by C. Domb and J. Lebowitz (Academic, London, 1983).

\bibitem{nubound}
J.T. Chayes {\it et al.} Phys. Rev. Lett. {\bf 57}, 2999 (1986).

\bibitem{maxent}
R.N. Silver, D.S. Sivia and J.E. Gubernatis, Phys. Rev. B {\bf 41}, 2380
(1990); M. Veki\'c and S.R. White, Phys. Rev. B {\bf 48}, 7643 (1993).

\bibitem{1dpapers}
S.K. Ma, C. Dasgupta, and C.K. Hu, Phys. Rev. Lett. {\bf 43}, 1434 (1979);
S.R. Bondeson and Z.G. Soos, Phys. Rev. B {\bf 22}, 1793 (1980);
J.E. Hirsch and J.V. Jos\'e, Phys. Rev. B {\bf 22}, 5339 (1980).

\bibitem{semicond}
R.N. Bhatt and P.A. Lee, Phys. Rev. Lett. {\bf 48}, 344 (1982).

\end{references}
\end{document}